\documentclass[conference]{IEEEtran}
\IEEEoverridecommandlockouts
\usepackage{cite}
\usepackage{url}
\usepackage{amsmath,amssymb,amsfonts}
\usepackage{algorithmic}
\usepackage{graphicx}
\usepackage{textcomp}
\usepackage{siunitx}
\usepackage[table]{xcolor}
\def\BibTeX{{\rm B\kern-.05em{\sc i\kern-.025em b}\kern-.08em
    T\kern-.1667em\lower.7ex\hbox{E}\kern-.125emX}}
\begin{document}

\title{AI-based Decision Support System for Heritage Aircraft Corrosion Prevention\\

\thanks{This research was carried out in the scope of the PROCRAFT project within the JPICH Conservation and Protection Call, supported by the following national funding organizations: Agence Nationale de la Recherche (ANR, France), Ministry of Universities and Research (MUR, Italy) and Ministry of Education, Youth and Sports (MEYS, Czech Republic). The research by M. Kuchař was partially supported by the Grant Agency of the Czech Technical University in Prague, grant No. SGS24/125/OHK2/3T/12. The work by T. Vyhlídal was partially supported by by the European Union under the project ROBOPROX, reg. no. CZ$.02.01.01/00/22\_008/0004590$.\\ \\
\textsf{\large Preprint submitted to IEEE}}
}

\author{\IEEEauthorblockN{1\textsuperscript{st} Michal Kuchař}
\IEEEauthorblockA{\textit{Dept. Instrumentation and Control Engineering} \\
\textit{Faculty of Mechanical Engineering}\\
\textit{Czech Technical University in Prague}\\
Prague, Czech Republic \\
michal.kuchar@fs.cvut.cz}
\and
\IEEEauthorblockN{2\textsuperscript{nd} Jaromír Fišer}
\IEEEauthorblockA{\textit{Dept. Instrumentation and Control Engineering} \\
\textit{Faculty of Mechanical Engineering}\\
\textit{Czech Technical University in Prague}\\
Prague, Czech Republic \\
jaromir.fiser@fs.cvut.cz}
\and
\IEEEauthorblockN{3\textsuperscript{rd} Cyril Oswald}
\IEEEauthorblockA{\textit{Dept. Instrumentation and Control Engineering} \\
\textit{Faculty of Mechanical Engineering}\\
\textit{Czech Technical University in Prague}\\
Prague, Czech Republic \\
cyril.oswald@fs.cvut.cz}
\and
\IEEEauthorblockN{4\textsuperscript{th} Tomáš Vyhlídal}
\IEEEauthorblockA{\textit{Dept. Instrumentation and Control Engineering} \\
\textit{Faculty of Mechanical Engineering}\\
\textit{Czech Technical University in Prague}\\
Prague, Czech Republic \\
tomas.vyhlidal@fs.cvut.cz}

}

\maketitle

\begin{abstract}
The paper presents a decision support system for the long-term preservation of aeronautical heritage exhibited/stored in sheltered sites. The aeronautical heritage is characterized by diverse materials of which this heritage is constituted. Heritage aircraft are made of ancient aluminum alloys, (ply)wood, and particularly fabrics. The decision support system (DSS) designed, starting from a conceptual model, is knowledge-based on degradation/corrosion mechanisms of prevailing materials of aeronautical heritage. In the case of historical aircraft wooden parts, this knowledge base is filled in by the damage function models developed within former European projects. Model-based corrosion prediction is implemented within the new DSS for ancient aluminum alloys. The novelty of this DSS consists of supporting multi-material heritage protection and tailoring to peculiarities of aircraft exhibition/storage hangars and the needs of aviation museums. The novel DSS is tested on WWII aircraft heritage exhibited in the Aviation Museum Kbely, Military History Institute Prague, Czech Republic.

\end{abstract}


\section{Introduction}
Nowadays, there are many organizations or institutes on national, e.g. \cite{NPU, BLfD, GPLA}, and international level, \cite{ICOM,ICCROM}, dealing with complex heritage safeguard. The complexity of safeguarding heritage consists of the simultaneous protection of the heritage displayed/exhibited or stored as well as historical buildings/sites themselves. The linkage between the architecture and the culture in the heritage is to be kept to safeguard both collections and historical architecture, \cite{Lucchi2}. Nevertheless, the "Modern Movement of Architecture" reshapes historical and museum architecture because of rational planning, natural light integration, innovative technologies, materials, etc.

An approach to safeguarding the heritage, see recent studies \cite{Lucchi1, Lucchi2, LaRussa}, relies on risk-based analysis for supporting a decision-making process on heritage conservation, energy efficiency, and visitor/staff comfort in historical buildings. This risk-based analysis is a multidisciplinary task involving physics, computer science, and information technologies (IT). A decision support system (DSS) becomes a tool for on-time and adequate safeguarding of the heritage in its complexity. The DSS regularly includes retrofit or refurbishment measures, and the natural goals of all the measures recommended/assigned by the DSS are an extension of the heritage lifetime and an improvement of the energy efficiency and the environmental performance of historic buildings/sites, as detailed in \cite{Lucchi1}. Another of these measures is the removal of the collection safety threats, for instance, textile carpets due to the resuspension; for more details on visitor-triggered resuspension, see \cite{Serfozo}. 

In \cite{Lucchi1}, SOBANE (screening, observation, analysis, expertise) methodology is applied to developing the DSS that utilizes identified heritage damage sensitivities and the environmental requirements for guaranteeing its conservation and management, considering the opposite needs of conservation and human comfort. Particularly, the environmental risks related to light conditions, air temperature, relative humidity, and pollution are mapped, and in the management, the risks with respect to storage conditions, exhibits, handling, and cleaning are in play. In \cite{Carlon}, the DSS called ArcheoRisk is developed to safeguard/rehabilitate archaeological sites in Venice Lagoon. Thereby, the archaeological risk evaluation and intervention selection is underpinned by the Geographical Information System (GIS) platform. In \cite{Viani}, E-Museum is developed to provide advanced services utilizing a Wireless Sensor Network (WSN) to collect real-time monitored data to facilitate the management of museum activities. While in \cite{Calvano}, the DSS is dedicated to optimizing visitor flows functionally to designing prospective exhibitions. In \cite{LaRussa}, enhanced usage of artificial intelligence (AI) enabled the achievement of decision-making tools based on machine learning via thermo-hygrometric and energy simulations of different scenarios for museums' use and management. Thereby, Historical Sentient - Building Information Modeling (HS-BIM) is integrated with this tool, assuming a synthetic behavior in the processing of the external and internal inputs of the historical building. However, AI-based risk analysis requires a large amount of data available for supervised training at best. Since these data are missing or lacking, the synthetic data are being obtained by numerical simulations instead; for more details, see \cite{LaRussa}. In \cite{Degrigny}, \textit{MiCorr} Decision Support System is developed within the MIFAC Metal project to diagnose the corrosion forms of heritage metals and to decide on conservation protocols (thus, to find treatment protocols). The decision-making process is based on the search for the closest match of observed metal corrosion with a database of corrosion models. Resources of basic corrosion models originate from \cite{Turgoose} for historical and archaeological lead, \cite{Robbiola} for archaeological copper alloys, and \cite{Dillmann} for archaeological iron. Furthermore, to get extensive and comprehensive corrosion models of historical and archaeological artifacts, the MIFAC Metal case studies, see \cite{Degrigny} and references therein, were applied to develop the \textit{MiCorr} DSS.

The conservation status of aeronautical heritage has been mapped years ago \cite{Degrigny21, Hallam, Mirambet, Brunet}. The heritage aircraft are composed of sandwich structures made of parts of aluminum alloys, wood (plywood. balsa), and fabrics (textile, canvas), in particular, \cite{Castanie}. As opposed to ancient wooden heritage collections, a pitfall of the aeronautical heritage protection consists in not restorable wooden aeronautical heritage, as a rule, see \cite{Sykora}, because of decaying this heritage into waterlogged archaeological wood, for more details on this kind of wood, see \cite{Pecoraro}. Another problem of the aeronautical heritage is, besides atmospheric corrosion, galvanic corrosion due to the presence of metals (particularly iron) in combination with aluminum alloys (duralumin, in particular), \cite{Carlos}. Aside from material degradation problems, the heritage aircraft hangars are deficient in equipment (e.g., HVAC) and insulation (shell buffering), as a rule. There are only a few studies coping with the protection of this heritage type \cite{Degrigny21, Hallam, Brunet2, Oswald}, but the earliest study on the protection of "ancient" aluminum alloys deployed in pre-war aircraft was presented already in 1934, \cite{Mutchler}. Of course, for modern aircraft, the particulate studies on aircraft protection from degradation have been done recently, \cite{Titakis, Diler, Macha}, and for modern aluminum materials, for instance, \cite{El-Mahdy}. Very recently, a comprehensive study on a preventive approach to aircraft heritage protection has been carried out in \cite{Kuchar}. Viewing the state-of-the-art above in the aeronautical heritage protection a novel DSS being developed and presented in this paper is tailored to the preventive approach from \cite{Kuchar}, with prospective intention of DSS's adaptability to different heritage sites.

\section{Design and implementation of decision support system} \label{sec:design_impl}
{First of all, a conceptual model is achieved in a DSS development to represent the knowledge base and to model the semantic data. The conceptual model is translated into a data structure and finalized into a visual outcome to work on a DSS and provide a linkage between reality and the data structure. The outcome is then a visual representation of the decision-making process (reasonings of actions to undertake) in favor of aeronautical heritage preservation as described in the DSS design below.

The DSS design for preserving aeronautical heritage is based on multiple-input and multiple-output (MIMO) decision tree model. The learning of a decision tree was chosen because we consider it desirable to design the DSS as data-driven due to the possible complexity of all input-output combinations rather than hardcoding all of the rules. However, the input-output combinations for learning were defined by experts or by literature knowledge at the first step. Adaptability is a cornerstone of this system's design. Upon introducing new input-output data, the decision tree requires retraining to assimilate the fresh data, thereby maintaining the accuracy and relevancy of its recommendations. The system’s input and output structure are presented in tabular forms (Tab. \ref{Tab:input_features}, Tab. \ref{Tab:possible_actions}). The DSS architecture in Fig. \ref{fig:dss_architecture} shows a comprehensive framework, integrating external and local environmental data sources, storing them in the database, and using state-of-the-art frameworks.

\begin{figure*}[ht]
\centerline{\includegraphics[width=1\textwidth]{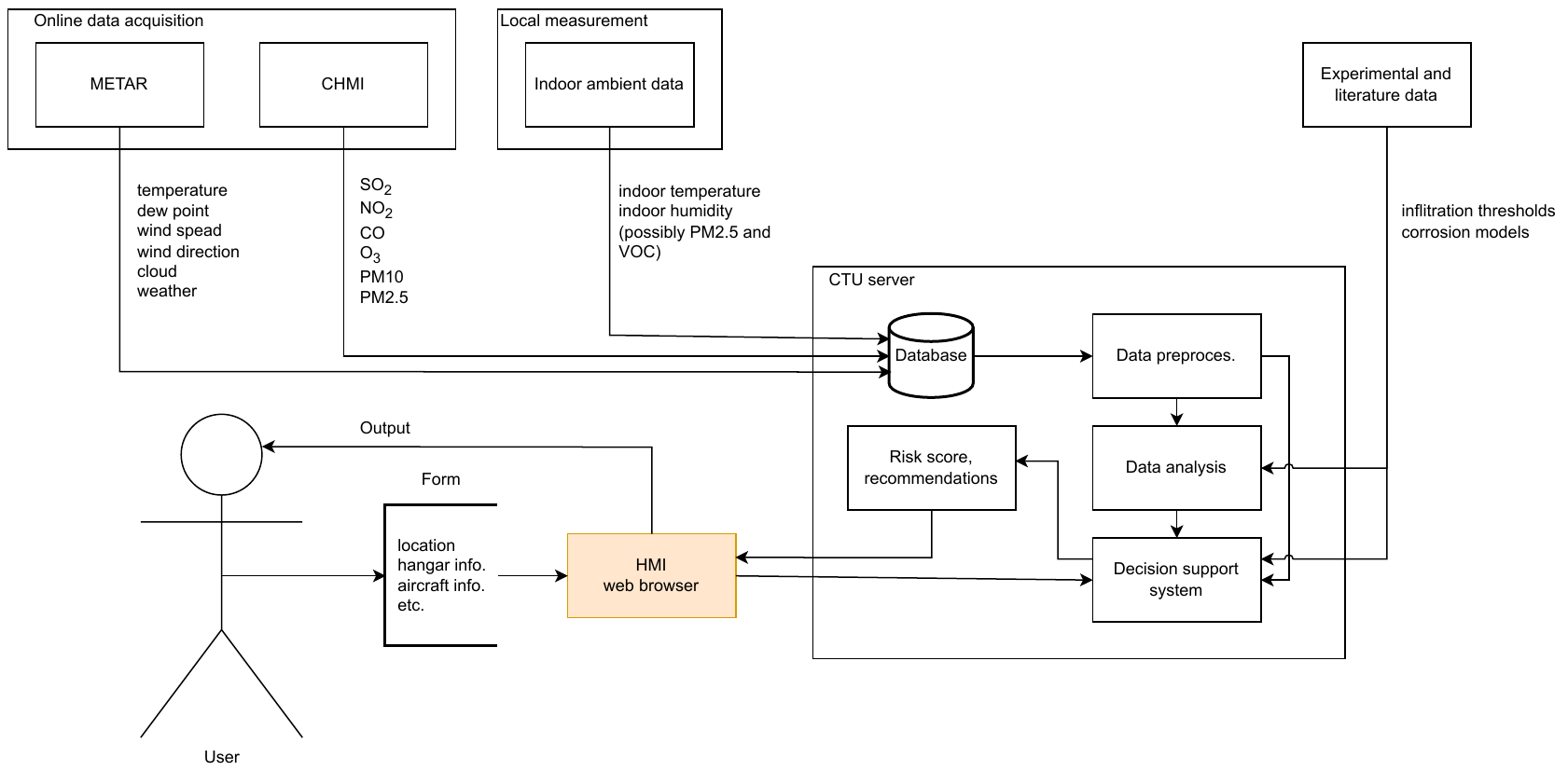}}
\caption{Architecture of decision support system \textit{SmartHangar}}
\label{fig:dss_architecture}
\end{figure*}

Online data acquisition is a crucial system component, utilizing METAR (Meteorological Terminal Air Report) reports \cite{METAR2025} to harvest real-time meteorological data, including air temperature, dew point, wind speed and direction, and general weather conditions. The Czech Hydrometeorological Institute (CHMI) data extends the system with localized pollutant concentrations, including $\unit{SO_2}$, $\unit{NO_2}$, $\unit{CO}$, $\unit{O_3}$, $\unit{PM_{10}}$, and $\unit{PM_{2.5}}$ levels. Complementing these datasets, local measurements from indoor ambient data provide crucial parameters such as indoor air temperature, humidity, and, if available, particulate matter and volatile organic compound (VOC) concentrations.

In tandem, the server employs a data-driven corrosion model obtained with the \emph{pycaret} framework, \cite{pycaret}, utilizing experimental data to evaluate corrosion risks. The model interprets ambient conditions, including air temperature, humidity, and pollutant levels, to calculate a risk score that quantifies the potential for corrosion. This approach uses the data-driven corrosion modeling from \cite{Kuchar}.

The fusion of experimental and literature data, predefined infiltration thresholds, and data-driven corrosion models empowers the DSS to deliver decisions catering to the nuanced needs of heritage aircraft preservation within the variable conditions of storage and exhibition environments.

The DSS is a complex software system that integrates state-of-the-art frameworks such as Scikit-learn, Pycaret, FastAPI to address the non-conventional use case of corrosion prediction in hangars

A Human-Machine Interface (HMI) is developed with Vue.js for user interaction. Through HMI, users will input data described in Tab. \ref{Tab:input_features}  using forms, which the DSS uses to generate corrosion risk estimation and actionable preservation recommendations.

\begin{table}[htbp]
\caption{Input features}
\begin{center}
\begin{tabular}{|c|c|}
\hline
\textbf{Ambient features} & \textbf{Data type}\\
\hline
\hline
Indoor air humidity [\unit{\percent}] & Time series \\
\hline
Outdoor air humidity[\unit{\percent}]  & Time series\\
\hline
Indoor air temperature [\unit{\celsius}]  & Time series\\
\hline
Outdoor air temperature [\unit{\celsius}]  & Time series\\
\hline
Indoor dew point [\unit{\celsius}] & Time series\\
\hline
Outdoor dew point [\unit{\celsius}] & Time series\\
\hline
Outdoor $\unit{SO_2}$ [$\unit{\micro\gram\per\cubic\m}$]  & Time series\\
\hline
Outdoor $\unit{PM_{2_{\cdotp}5}}$ [$\unit{\micro\gram\per\cubic\m}$]  & Time series\\
\hline
Outdoor $\unit{PM_{10}}$ [$\unit{\micro\gram\per\cubic\m}$]  & Time series\\
\hline
Wind speed [\unit{\meter\per\second}]  & Time series\\
\hline
Air exchange rate [\unit{\per\hour}] & Time series\\
\hline
\hline
\textbf{Calculated features} & \textbf{Data type}\\
\hline
\hline
Indoor $\unit{SO_2}$ [$\unit{\micro\gram\per\cubic\m}$]  & Time series \\
\hline
Time of wetness [\unit{\hour}]  & Numeric \\
\hline
\textbf{Hangar information} & \textbf{Data type}\\
\hline
\hline
Location near sea \{yes, no\} & Binary \\
\hline
AC installed \{yes, no\} & Binary \\
\hline
Heating installed \{yes, no\} & Binary \\
\hline
Filters installed \{yes, no\} & Binary\\
\hline
Insulation installed \{yes, no\} & Binary\\
\hline
Barriers installed \{yes, no\} & Binary \\
\hline
Carpets installed \{yes, no\} & Binary \\
\hline
Walls material \{wood, steel, concrete\} & Categorical\\
\hline
Walls area [\unit{\meter\squared}] & Numeric\\
\hline
Roof material \{wood, steel, concrete\} & Categorical\\
\hline
Roof area [\unit{\meter\squared}] & Numeric\\
\hline
Floor material \{wood, steel, concrete\} & Categorical\\
\hline
Floor area [\unit{\meter\squared}] & Numeric\\
\hline
Exhibition area [\unit{\meter\squared}] & Numeric \\
\hline
Hangar volume [\unit{\cubic\meter}] & Numeric\\
\hline
\end{tabular}
\label{Tab:input_features}
\end{center}
\end{table}

\begin{table}[htbp]
\caption{Possible actions}
\begin{center}
\begin{tabular}{|c|}
\hline
\textbf{Possible actions} \\
\hline
\hline
Increase or decrease air exchange rate \\
\hline
Start or stop heating \\
\hline
Start or stop AC \\
\hline
Increase or decrease number of people in the hangar \\
\hline
Change the ratio between exhibition area and hangar volume \\
\hline
\hline
\textbf{Refurbishment} \\
\hline
\hline
Install filters (HEPA, carbon etc.) \\
\hline
Install AC \\
\hline
Install heating \\
\hline
Install insulation \\
\hline
Install barriers \\
\hline
Uninstall carpets \\
\hline
\end{tabular}
\label{Tab:possible_actions}
\end{center}
\end{table}

\subsection{Used technologies}
As mentioned before, the development of our DSS, called \textit{SmartHangar}, incorporates multiple frameworks. Below, we detail and justify the selection of each tool within our technology stack.

\emph{PostgreSQL}: Our choice for the database management system is PostgreSQL due to its strong reputation for data integrity and its support for advanced data types. Its robust feature set, including complex queries, extensible indexing, and concurrency without read locks, makes it an ideal choice for managing the complex and relational data structures that DSS \textit{SmartHangar} requires.

\emph{FastAPI}: FastAPI is employed to create a modern, fast (high-performance) web framework for building APIs. It is selected for its speed, ease of use, and robustness. Its ability to handle asynchronous requests ensures our system can manage a high volume of requests efficiently, which is crucial for real-time data processing and responsiveness.

\emph{PyCaret}: In building our predictive models, PyCaret offers a low-code machine learning library in Python that automates model training and evaluation. It streamlines the workflow, which enables us to iterate rapidly through different models and preprocessing techniques. Its simplicity and integration with other libraries offer the flexibility needed in a research-oriented development environment.

\emph{Scikit-learn}: The backbone of our machine learning capability is formed by Scikit-learn. This open-source tool is used to implement the MIMO decision tree model. The library is renowned for its broad range of algorithms, consistent programming interface, and comprehensive documentation, which significantly expedite the development process.

\emph{Vue.js} for HMI Development: The Human-Machine Interface (HMI) of our system is now being developed using Vue.js, a progressive JavaScript framework. Vue.js is specifically chosen for its simplicity, detailed documentation, and flexibility. It allows for building a reactive, component-driven UI for our DSS, ensuring an intuitive and smooth user experience. Its modular architecture facilitates easy maintenance and the iterative development of the HMI as user requirements evolve.

The integration of these technologies results in a DSS that is functional, adaptable, and easy to update as requirements evolve. The chosen tools enable efficient corrosion prediction, supporting the preservation of aeronautical heritage.

\subsection{Preprocessing}
Before analysis, resampling was applied to standardize the temporal resolution across disparate data sources. Linear interpolation was used to fill gaps between existing observations, ensuring all datasets were aligned to a uniform date-time format. This step was essential for synchronizing datasets with varying acquisition frequencies and enabling coherent aggregation.

After resampling, moving average (MA) filters were applied to smooth the time series data, reducing short-term fluctuations and highlighting underlying trends. The sliding window size of the MA filter was a key parameter influencing the accuracy of the corrosion model. The optimal window length was determined through a grid search over values ranging from 1 to 168 hours.

\subsection{Calculated features}
The preprocessing pipeline also includes the computation of derived features. One key feature is the Time of Wetness, defined as the total number of hours when relative humidity (RH) exceeds 80 \% and temperature remains above 0°C. This parameter serves as an indicator of corrosion risk in the hangar.

Another important feature is indoor pollution concentration, estimated using outdoor pollution levels, the air exchange rate (\textit{n}), and the materials of both the hangar and stored artifacts. This calculation accounts for pollutant infiltration and interaction with the indoor environment, providing insight into conservation conditions.

These computed features (see Tab. \ref{Tab:input_features}) enhance the dataset and directly inform the DSS’s recommendations for corrosion prevention.

\subsection{Risk evaluation}
As mentioned in section \ref{sec:design_impl} and as it is shown in Fig. \ref{fig:raw_data}, the first fundamental tool for corrosion risk evaluation is a data-driven corrosion model, which evaluates mainly ambient data, including dew point (DP) temperature. However, we also use the ISO 9223 standard for corrosivity risk evaluation as described in \cite{Kuchar2}, which considers the Time of Wetness and pollution infiltration. Finally, additional rules, such as proximity to the sea or missing building insulation, are used. These parameters are evaluated within the DSS only as input for the decision tree.

\section{Use case}

The use case of DSS \textit{SmartHangar} is applied to the Aviation Museum Kbely. Measured ambient data are visualized in Fig. \ref{fig:raw_data} and pollution data in Fig. \ref{fig:pollution}. These data were published in \cite{Kuchar}. The rest of input data are described in Tab. \ref{Tab:input_features_use_case}.

As visible in Fig. \ref{fig:raw_data}, the output of the data-driven corrosion model predicts corrosion risk, which corresponds with high humidity and temperature jump from freezing to temperatures above zero.

\begin{figure}[h]
\centering
\includegraphics[width=\linewidth]{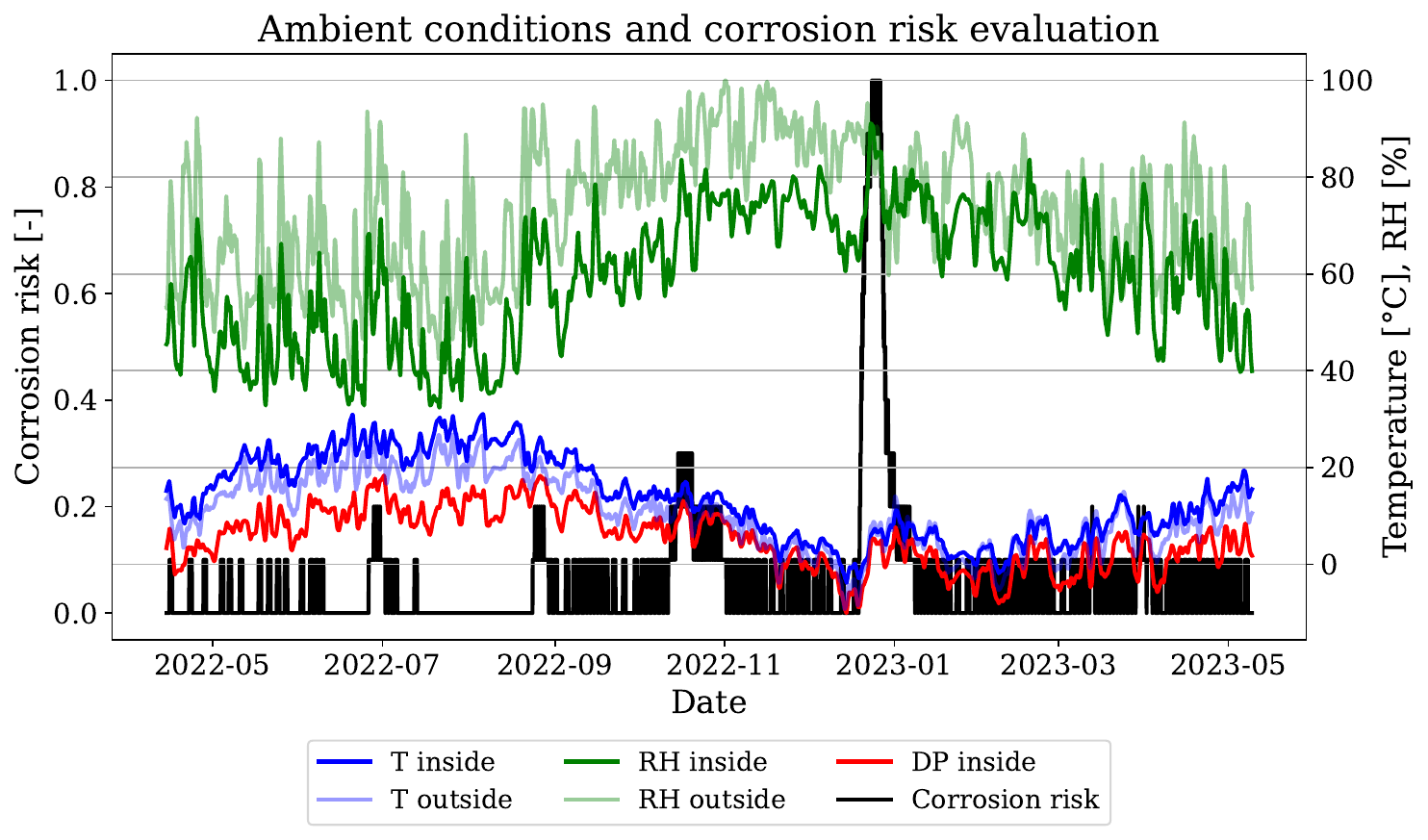}
\caption{Measured one-year data and corrosion risk score evaluated.}
\label{fig:raw_data}
\end{figure}  

\begin{figure}[h]
\centering
\includegraphics[width=\linewidth]{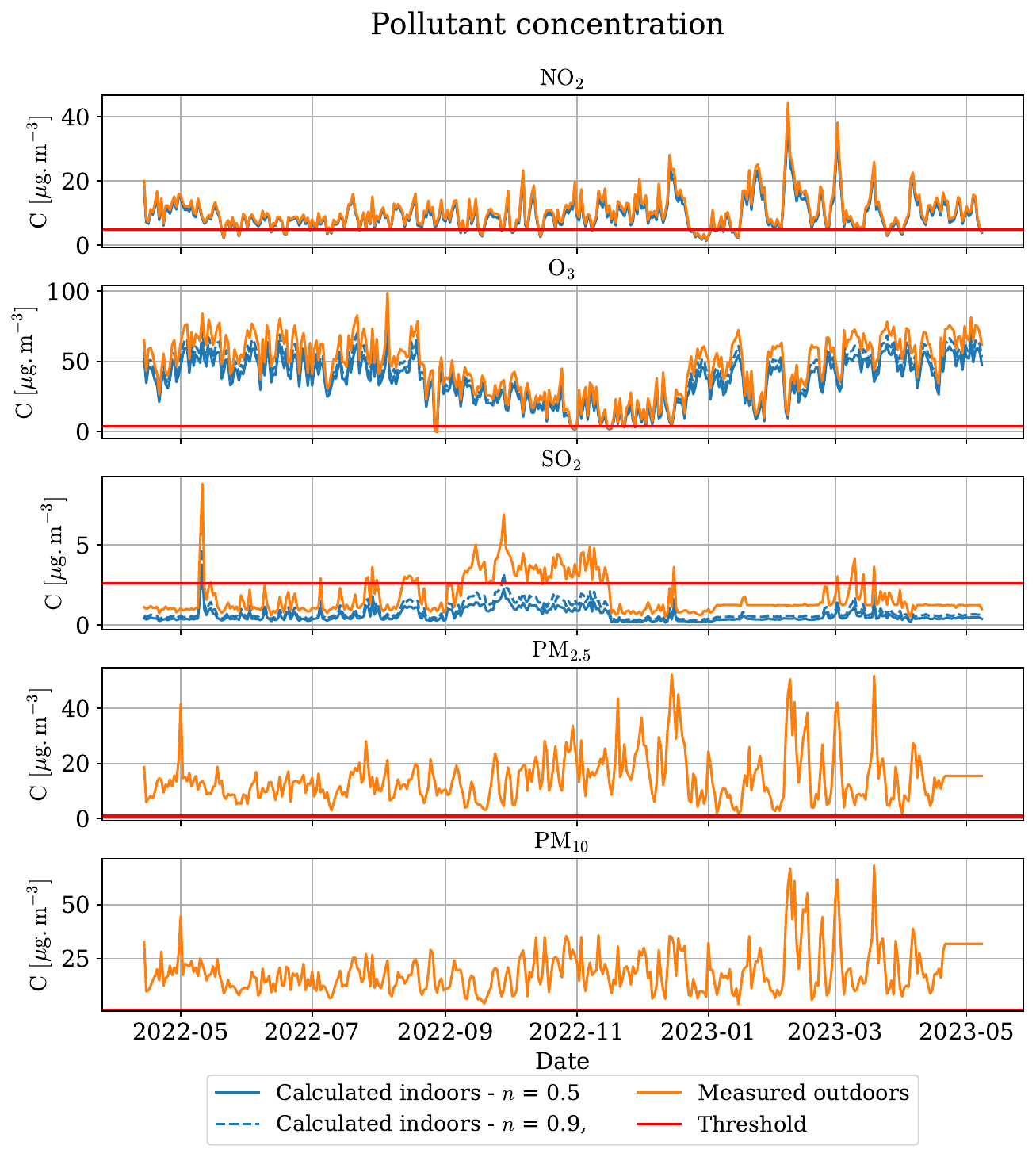}
\caption{One-year pollution data measured from weather stations in Prague-Holesovice and Prague-Riegrovy sady and indoor pollution calculated for min-max air exchange rate - Daily averages \cite{Kuchar}.}
\label{fig:pollution}
\end{figure}

\begin{table}[htbp]
\caption{Input features - use case}
\begin{center}
\begin{tabular}{|c|c|}
\hline
\textbf{Calculated features} & \textbf{Value}\\
\hline
\hline
Time of wetness [\unit{\hour}]  & 60.6 \\
\hline
\textbf{Hangar information} & \textbf{Value}\\
\hline
\hline
Location near sea \{yes, no\} & No \\
\hline
AC installed \{yes, no\} & No \\
\hline
Heating installed \{yes, no\} & No \\
\hline
Filters installed \{yes, no\} & No\\
\hline
Insulation installed \{yes, no\} & No\\
\hline
Barriers installed \{yes, no\} & No \\
\hline
Carpets installed \{yes, no\} & Yes \\
\hline
Walls material \{wood, steel, concrete\} & Wood\\
\hline
Walls area [\unit{\meter\squared}] & 1004.8\\
\hline
Roof material \{wood, steel, concrete\} & Steel\\
\hline
Roof area [\unit{\meter\squared}] & 985.6 \\
\hline
Floor material \{wood, steel, concrete\} & Concrete\\
\hline
Floor area [\unit{\meter\squared}] & 985.6\\
\hline
Exhibition area [\unit{\meter\squared}] & 985.6 \\
\hline
Hangar volume [\unit{\cubic\meter}] & 7884.8\\
\hline
\end{tabular}
\label{Tab:input_features_use_case}
\end{center}
\end{table}

The output of DSS \textit{SmartHangar} is shown in Tab. \ref{Tab:actions_use_case}.

\begin{table}[htbp]
\caption{Output of DSS \textit{SmartHangar}}
\begin{center}
\begin{tabular}{|c|c|}
\hline
\textbf{Actions} & Output\\
\hline
\hline
Increase or decrease air exchange rate & No action\\
\hline
Start or stop heating & No action\\
\hline
Start or stop AC & No action\\
\hline
Increase or decrease number of people in the hangar & No action\\
\hline
Change the ratio between exhibition area and hangar volume & No action\\
\hline
\hline
\textbf{Refurbishment} & Output \\
\hline
\hline
Install filters (HEPA, carbon etc.) & No action\\
\hline
Install AC & No action\\
\hline
Install heating & \cellcolor{green} Yes\\
\hline
Install insulation & \cellcolor{green} Yes\\
\hline
Install barriers & No action \\
\hline
Uninstall carpets & \cellcolor{green} Yes \\
\hline
\end{tabular}
\label{Tab:actions_use_case}
\end{center}
\end{table}

The corrosivity category by ISO 9223 standard is $\unit{C_2}$ - low (indoor climates without microclimate control like storages).

A partial demonstration of the DSS results in a web browser is shown in Fig. \ref{fig:screenshot}.

\begin{figure}[h]
\centering
\includegraphics[width=\linewidth]{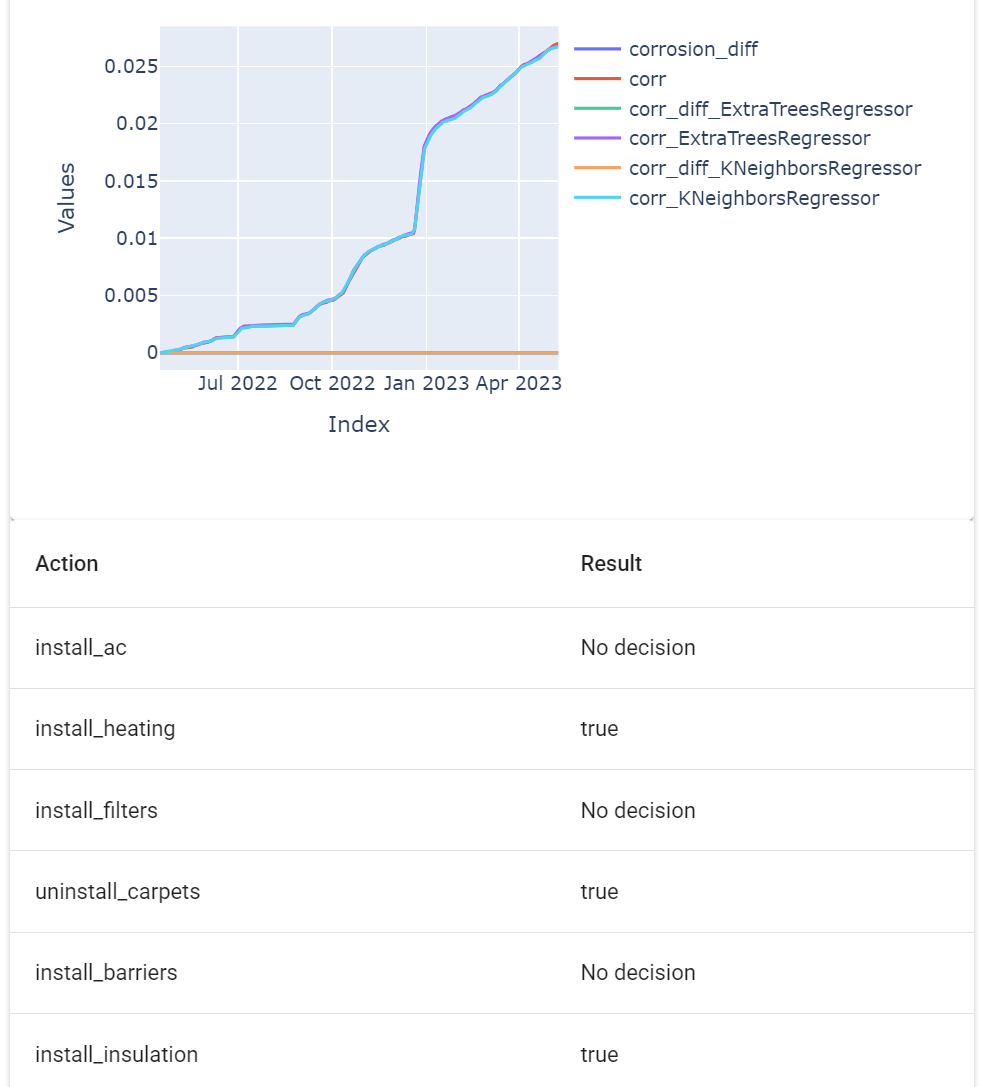}
\caption{Screenshot of the results part in HMI of \textit{SmartHangar}.}
\label{fig:screenshot}
\end{figure}  

\section*{Acknowledgment}

The authors thank Miroslav Khol, affiliated with Military History Institute Prague, Czech Republic, for his valuable insights into aeronautical heritage handling and management, for which we are very grateful.

\section*{Conclusions}

A novel decision support system is developed to preserve aeronautical heritage sheltered in hangars. After processing all the input features (external and internal) and applying the decision tree model, the output features (actions and risk scores) are obtained, providing recommendations for aeronautical heritage preservation. The proposed DSS's use case demonstrates its function in favor of aircraft heritage protection. Other use cases are needed to test the designed DSS on aviation museums located in different geographical locations.

\end{document}